\documentclass[conference]{IEEEtran}
\IEEEoverridecommandlockouts
\usepackage{cite}
\usepackage{amsmath,amssymb,amsfonts}
\usepackage{algorithmic}
\usepackage{graphicx}
\usepackage{textcomp}
\usepackage{xcolor}
\usepackage{amsmath}
\usepackage{balance}
\usepackage{todonotes}
\usepackage{url}
\usepackage[]{footmisc}
\usepackage{enumitem}
\def\BibTeX{{\rm B\kern-.05em{\sc i\kern-.025em b}\kern-.08em
    T\kern-.1667em\lower.7ex\hbox{E}\kern-.125emX}}

\begin{document}

\title{DHP Framework: Digital Health Passports \\Using Blockchain \\ \LARGE{Use case on international tourism during the COVID-19 pandemic} 
{\footnotesize \textsuperscript{}}
\thanks{}
}

\author{\IEEEauthorblockN{Constantinos Marios Angelopoulos\IEEEauthorrefmark{1},
Amalia Damianou and
Vasilis Katos}
\IEEEauthorblockA{Bournemouth University, Poole, UK
}
Email: \{mangelopoulos, adamianou, vkatos\}@Bournemouth.ac.uk}

\maketitle

\begin{abstract}
In order to contain the COVID-19 pandemic, several countries enforced extended social distancing measures for several weeks, effectively pausing the majority of economic activities. In an effort to resume economic activity safely, several \textit{Digital Contact Tracing} applications and protocols have been introduced with success. However, DCT is a \textit{reactive} method, as it aims to break \textit{existing} chains of disease transmission in a population. Therefore DCT is not suitable for \textit{proactively} preventing the spread of a disease; an approach that relevant to certain use cases, such as international tourism, where individuals travel across borders. In this work, we first identify the limitations characterising DCT related to privacy issues, unwillingness of the public to use DCT mobile apps due to privacy concerns, lack of interoperability among different DCT applications and protocols, and the assumption that there is limited, local mobility in the population. We then discuss the concept of a \textit{Health Passport} as a means of verifying that individuals are disease risk-free and how it could be used to resume the international tourism sector. Following, we present the DHP Framework that uses a private blockchain and Proof of Authority for issuing \textit{Digital Health Passports}. The framework provides a distributed infrastructure supporting the issuance of DHPs by foreign health systems and their verification by relevant stakeholders, such as airline companies and border control authorities. We discuss the attributes of the system in terms of its usability and performance, security and privacy. Finally, we conclude by identifying future extensions of our work on formal security and privacy properties that need to be rigorously guaranteed via appropriate security protocols. 

\end{abstract}

\section{Introduction}

The world has always been experiencing endemic and pandemic outbreaks and their potential impact as global threats is well established. However, in recent years the frequency and likelihood of such outbreaks as well as their collateral economic damage is continuously growing  \cite{wef}. Related research indicates that this is due to several factors favouring the spread of zoonotic diseases, such as the ever increasing urbanization and deforestation, the ongoing climate crisis and the increasingly globalized and connected world economy \cite{gordon2016increase}. Already, in the first two decades of the 21\textsuperscript{th} century there have been several global outbreaks, including those caused by the SARS-CoV (2003), H1N1 (2009), MERS-CoV (2012), Ebola (2014), Zika (2015) and SARS-CoV-2 (2020) viruses. 

Particularly in the case of SARS-CoV-2, the implications of the associated disease, COVID-19, have been globally profound. Initially reported in Wuhan, People's Republic of China in December 2019, the virus has practically spread to all countries in the world in the following months. By May 2020, the World Health Organisation (WHO) reported nearly four million COVID-19 cases and more than two hundred sixty thousand related deaths worldwide \cite{world2020coronavirus}. The spread was rapid and caused, in a short period of time, a significantly increased number of patients requiring admission to hospitals and intensive care units, resulting to some  health systems to be overwhelmed \cite{Emanuel2020}. 

The response by governments in trying to initially reduce the infection rate and eventually contain the pandemic was to adopt extended social distancing measures and in several cases to impose total lockdowns. The measures were in place for a typical period of three months before starting to be partially lifted subject to the rate of infection being adequately low \cite{alvarez2020simple}. 

As a result, the economic activity in almost all sectors paused, except for key activities related to health and safety services, and distribution of food and essential goods. The economic repercussions were severe; indicatively, the World Bank predicted in a preliminary report in April 2020 a 4\% contraction of the global GDP under an amplified global pandemic \cite{worldbank}; the World Trade Organisation expected global trade to fall up to 32\% in 2020 \cite{Fernandes2020}; reports estimated the drop in the USA GDP to be as high as 11\%\cite{NBERw26983}; and the European Commission expected a record 7\% contraction of the EU economy \cite{euspring2020}. For countries whose economies have great exposure on international tourism, such as Mexico, Spain and Greece, contraction would be even greater, reaching or exceeding 15\%\cite{Fernandes2020}. A positive linear correlation was found between the duration of the measures and the negative impact on economy. Therefore, the need to fine tune the balance between reducing the negative economic impact and containing the pandemic necessitated for innovative and effective lockdown exit strategies.

In laying down such exit strategies, the WHO recommendations on monitoring and controlling a pandemic need to be considered: rapid diagnosis and immediate isolation of cases; rigorous tracking;
and precautionary self-isolation of recent social contacts \cite{whoguides}. In the case of SARS-CoV-2 lack of accurate rapid tests (since SARS-CoV-2 was new) means that the testing capacity and throughput of health systems was limited. Therefore much focus was placed on rigorous contact tracing of newly identified COVID-19 patients. This approach aimed to increase the effectiveness of diagnostic testing by promptly identifying asymptomatic patients and pre-emptively removing them from the general population via self-isolation. Empirically, it was proven that countries that adopted this strategy were successful in containing the pandemic in its early stages\cite{Salath2020}. 

However, the testing-tracing-isolating strategy remains a \emph{reactive} strategy. It only allows for a "trial-and-monitor" approach in which social distancing measures are tightened and loosened in an iterative manner while monitoring the infection rate of a population. This strategy does not allow for a safe and total restart of the local and global economy. For instance, it does not allow for the informed reopening of hospitality businesses (restaurants, entertainment venues, etc) or the full return to work of office employees. At a global scale, reactive strategies do not allow for international travels to resume and therefore economies that have significant exposure to international tourism remain extremely vulnerable.

In this context, some governments have suggested that the detection of antibodies to the SARS-CoV-2 could serve as the basis for an "\textit{immunity passport}" or \textit{risk-free certificate} that would enable individuals to travel or to return to work assuming that they are protected against re-infection. However, by the time of writing, there is no evidence that people who have recovered from COVID-19 and have antibodies are protected from a second infection \cite{whopassport}.

\textbf{Our contribution.} There is a need for a framework that will allow for \textit{continuous}, \textit{scalable}, \textit{reliable}, \textit{secure}, \textit{trustworthy} and \textit{timely} sharing of information across different health systems and countries on whether individuals are infected with SARS-CoV-2. Our contribution is summarised as follows:
\begin{itemize}
\item We analyse Digital Contact Tracing as a method of monitoring and preventing the spread of a contagious disease such as COVID-19. We identify (a) its dependency upon public perception and adoption (mainly due to privacy concerns); (b) its reactive nature, as it assumes that the disease is already present in the population; and (c) the problem of interoperability among different DCT applications and protocols. These limitations make DCT not fit for purpose in use cases, such as international tourism, where individuals need to travel over long distances and across borders.
\item Motivated by the dependency of several national economies (such as those of Mexico, Portugal, and Greece) on international tourism, and the relevant announcements by the World Health Organisation and several national governments, we discuss the concept of a Digital Health Passport. We outline a corresponding use case and define the relevant functional and technical requirements.
\item We present \textit{DHP Framework}; a scalable, distributed and secure framework for issuing Digital Health Passports (DHP). The framework provisions the secure issuance of digitally signed DHPs by local health facilities to individuals following medical tests. Each health facility announces the DHPs to its corresponding national health authority, that proceeds with registering them on a private blockchain. The health authorities of multiple countries form part of the DHP Blockchain consortium, which uses Proof of Authority as a consensus mechanism. The consortium also includes members with read-only access on the blockchain, such as airline companies and border control authorities. These members are able to check the validity of the health passport only for the travellers they are concerned and in a way that respects their privacy and their right to be forgotten. 
\item We analyse the presented framework with respect to its functionality, as well as its security and privacy aspects. We conclude the paper, by identifying future extensions of our work on formal security and privacy properties that need to be rigorously guaranteed via appropriate security protocols.
\end{itemize}

The paper is organised as follows. In Section \ref{rw}, we provide the background and cover the related work on strategies for controlling epidemics, we discuss Digital Contact Tracing and provide its limitations, and we provide a brief introduction to blockchain technologies and key relevant concepts. In Section \ref{usecase}, we discuss the concept of Digital Health Passports and their application in international tourism during an ongoing global pandemic. In Section \ref{framework} we present in detail the DHP Blockchain framework; the involved actors, data structures and procedures. In Section \ref{analysis}, we discuss the attributes of the system in terms of its usability and performance, security and privacy. Finally, in Section \ref{future}, we conclude by identifying future extensions of our work on formal security and privacy properties that need to be rigorously guaranteed via appropriate security protocols.  

\section{Background and Related Work}
\label{rw}
\subsection{Epidemic control strategies and their impact}
The modern globalized world economy heavily relies on international travels and global trade. As a result, widespread outbreaks, such as the COVID-19 pandemic caused by the SARS-CoV-2 virus, are very hard to contain in a single country or region mainly due to the increased connectedness of modern societies. In this context, WHO has identified the best strategy for monitoring and controlling a pandemic in a combination of large-scale testing, rigorous contact tracing, isolation and quarantine, in parallel with moderate (e.g., South Korea) or strong (e.g., China) social-distancing measures \cite{whoguides, Gilbert2020}. The strategy aims at first \textit{flattening the infection curve}; i.e. in reducing the rate at which new patients contract the virus, such that local health systems are given enough time to upscale their capacity and not be overwhelmed \cite{Anderson2020}. 

In principle, the strategy works by breaking the \textit{chain of transmission} and by \textit{promptly identifying clusters of infection} in the population, which subsequently are isolated from the rest of the population \cite{Fong2020,Kissler2020}. In practice, however, how the strategy is implemented depends on several externalities, such as: how quickly and how strictly the measures will be applied by local authorities \cite{Ganem2020}; how much the local population will adhere to the measures \cite{Painter2020, Allcott2020, Makridis2020} and what is the level of trust to authorities \cite{Blair2017, Trapido2019}; and the characteristics of the virus (incubation period, mode of transmission, etc.) \cite{Li2020, Lipsitch2003}.

In the case of SARS-CoV-2, social distancing measures (whether strict or moderate) were maintained for an average period of three months prior to being relaxed \cite{alvarez2020simple}. Within this period, the grand majority of the economic activities were halted at national and regional levels, while there were severe disruptions at a global level. The World Bank predicted in a preliminary report in April 2020 a 4\% contraction of the global GDP under an amplified global pandemic \cite{worldbank}; reports estimated the drop in the USA GDP to be as high as 11\%\cite{NBERw26983}; and the European Commission expected a record 7\% contraction of the EU economy \cite{EU}.
    
\subsection{Digital Contact Tracing and its limitations}
In the context of the testing-tracing-isolating strategy, several countries employed the use of digital technology in order to perform rigorous contact tracing of newly identified COVID-19 patients. In \cite{Ferretti2020} authors showed that \textit{Digital Contact Tracing} (DCT) - i.e., the use of smartphones to trace and notify recent contacts of a newly identified patient - and immediate notification are sufficient to stop the epidemic if the corresponding mobile applications are used by a sufficiently high proportion of the population. Due to the use of personal devices, DCT applications pose several issues with regards to the personal privacy of their users \cite{Tang2020, Cho2020} and as a result public concerns were raised. 

Several DCT protocols and applications have been developed to address the  privacy issues, such as PACT \cite{Chan2020}, CONTAIN \cite{Hekmati2020}, TraceSecure \cite{Bell2020} and the NHS COVID-19 app \cite{nhscovid} in the UK. Passive DCT methods include the use of GPS data and Bluetooth beacons while participatory ones include scanning of QR codes by the user \cite{Li2020a}. DCT methods are also distinguished in centralised and decentralised ones \cite{Li2020a}, while there have been efforts for hybrid approaches \cite{beskorovajnov2020contra}. Despite the privacy considerations, DCT methods, have been proven effective in helping break new transmission chains in a population. 

However, we identify significant limitations. 
\begin{itemize}
\item Firstly, \textit{DCT is dependant upon the perception of and its adoption by the general public}. As public awareness on privacy issues - particularly in the context of digital systems - is growing globally, citizens are cautious and perhaps reluctant to use or voluntarily disclose data on their activities. This does not make the method reliable for monitoring and controlling pandemics in the general case. 
\item Secondly, \textit{DCT methods are \textit{reactive} in nature}. While they help identify and break new transmission chains of infection, they do not provide a means for protecting against their initiation. On the contrary, the method helps identify potential future cases given that the disease is already present in the population. 
\item Thirdly, in the general case, \textit{different DCT applications are not interoperable with each other}. Apart from not allowing for seamless exchange of data, lack of interoperability also makes DCT unsuitable as a method for international travels as a single DCT application is expected to be deployed in a specific country and as such assumes a moderate to low mobility of the population.
\end{itemize}

\subsection{Blockchain: a brief introduction}
Blockchain is a distributed ledger technology where data (commonly refered to as transactions) are registered in blocks that are linked to each other using cryptography, thus forming a chain \cite{drescher2017blockchain}. New transactions are registered in new blocks that are created and appended to the chain by special actors in the network; these are referred to as miners, validators or authorities depending on the design of the network. Consensus mechanisms are used in order for the members of the consortium to validate and agree on the current state of the blockchain. Indicative consensus mechanisms include Proof of Work - where the validation of a new block requires solving computationally hard puzzles; Proof of Stake - where the validation requires the monetary commitment of the validator; and Proof of Authority - where the validators are considered trust-worthy, honest and their identity is well-known. Blockchain consortia can also be distinguished to public, private and permissioned with respect to access rights. Public blockchains are open access and any member of the network can join and participate in the core activities of the network. Private blockchains allow only selected entry of verified participants, while permissioned blockchains allow a mixed approach where members can join after suitable verification and can be assigned roles with different rights in the network.

\section{Digital Health Passports for Proactive Pandemic Containment}
\label{usecase}
\subsection{Motivation: International tourism and the case of Greece }
Among the most severely affected economic sectors by the COVID-19 pandemic have been global trade and international tourism. The World Trade Organisation expected global trade to fall up to 32\% in 2020 due to te COVID-19 pandemic \cite{Fernandes2020}. The effects of COVID-19 in the tourism, hospitality and recreation sectors have been unprecedented. In the accommodation and lodging sectors, quarterly revenues went down 75\%, while travel agents saw a slowdown in bookings of 50\% in March of 2020 \cite{worldbank}. As of mid-March 2020, international travel was ground to a halt, with the World Travel and Tourism Council (WTTC) estimating that global travel would decline at least 25\% in 2020 anticipating a loss of up to USD2.1 trillions in 2020. The International Air Transport Association (IATA) projected a resulting revenue loss for airlines of USD252 billion \cite{iata}. 

These developments represented a major risk to countries whose economies have significant exposure to international tourism. For instance, in Greece the impact of the crisis was forecast to be large due to the importance of the hospitality sector in economy (the sector represents more than 20\% of the national GDP) and the high share of micro enterprises, which are vulnerable \cite{euspring2020}. The pandemic came at a time when the Greek economy was recovering after ten years of austerity measures following the financial crisis of 2008. 

After having imposed strict social distancing and self-isolation measures promptly, Greece was successful in managing the pandemic. As a result, in May 2020, Greek authorities partially lifted the measures with the aim of opening the country for the upcoming summer tourist season. Nevertheless, how to effectively safeguard local population from subsequent pandemic waves while at the same time welcoming and catering for international tourists remained an open question. 

In an effort to address the issue, the concept of some sort of "\textit{immunity passport}" or \textit{risk-free certificate} has been considered \cite{whopassport}. The idea, as announced by the Greek Governement, was for international travellers to be tested for COVID-19 at least 72 hours prior to their departure and to be allowed to enter the country subject to them having tested negative to the virus \cite{grcovid}. Contrary to the \textit{reactive} strategy of testing-tracing-isolating, this was a \textit{proactive} strategy that sought to obstruct infected (and potentially asymptomatic) individuals from entering the country.  However, the announced approach posed several issues. In particular:

\begin{itemize}
\item Firstly, it required international tourists to take the diagnostic test several days before their flight, thus not addressing the possibility of contracting the virus in the following days. 
\item Secondly, the details on who the certifying authority would be (e.g. local private clinic or a certified health unit) were not specified. A standardised approach is needed that will be able to address discrepancies among different health systems, diagnostic methods, as well as for attribution purposes (e.g. in the case of fraudulent documents issued illegally).
\item Thirdly, the form of the certificate was not specified. If the certificate is in the form of a printed document, then it is susceptible to forgery. A digital form would provide guarantees in this respect, but this approach raises the following point.
\item In the case of a digital certificate, the challenge of how will information be shared seamlessly and timely among different stakeholders (e.g. airlines, border authorities, etc) across different countries while respecting the privacy of the travellers needs to be addressed.    
\end{itemize}

\subsection{The use case on international tourism}

We consider the case of international tourism, where people travel by airplane across different countries. During a pandemic, such as that of COVID-19, we assume that \textit{hygiene protocols} are enforced by countries that specify a series of entry conditions for international travellers (similar to those for travel visas). In this study, motivated by the announcements of the Greek government in May 2020 \cite{grcovid}, we assume that the hygiene protocols require the travellers  to take a diagnostic test at a given time prior to the date of travel and that the test needs to be conducted using a specific method. The latter is motivated by the fact that the RT-qPCR method (the main diagnostic method for COVID-19 by May 2020) can only achieve a sensitivity between 50\% and 79\% \cite{li2020molecular} and therefore, at the time of writing, additional methods are being researched.

\begin{figure*}[]
\centering
\includegraphics[width=\textwidth]{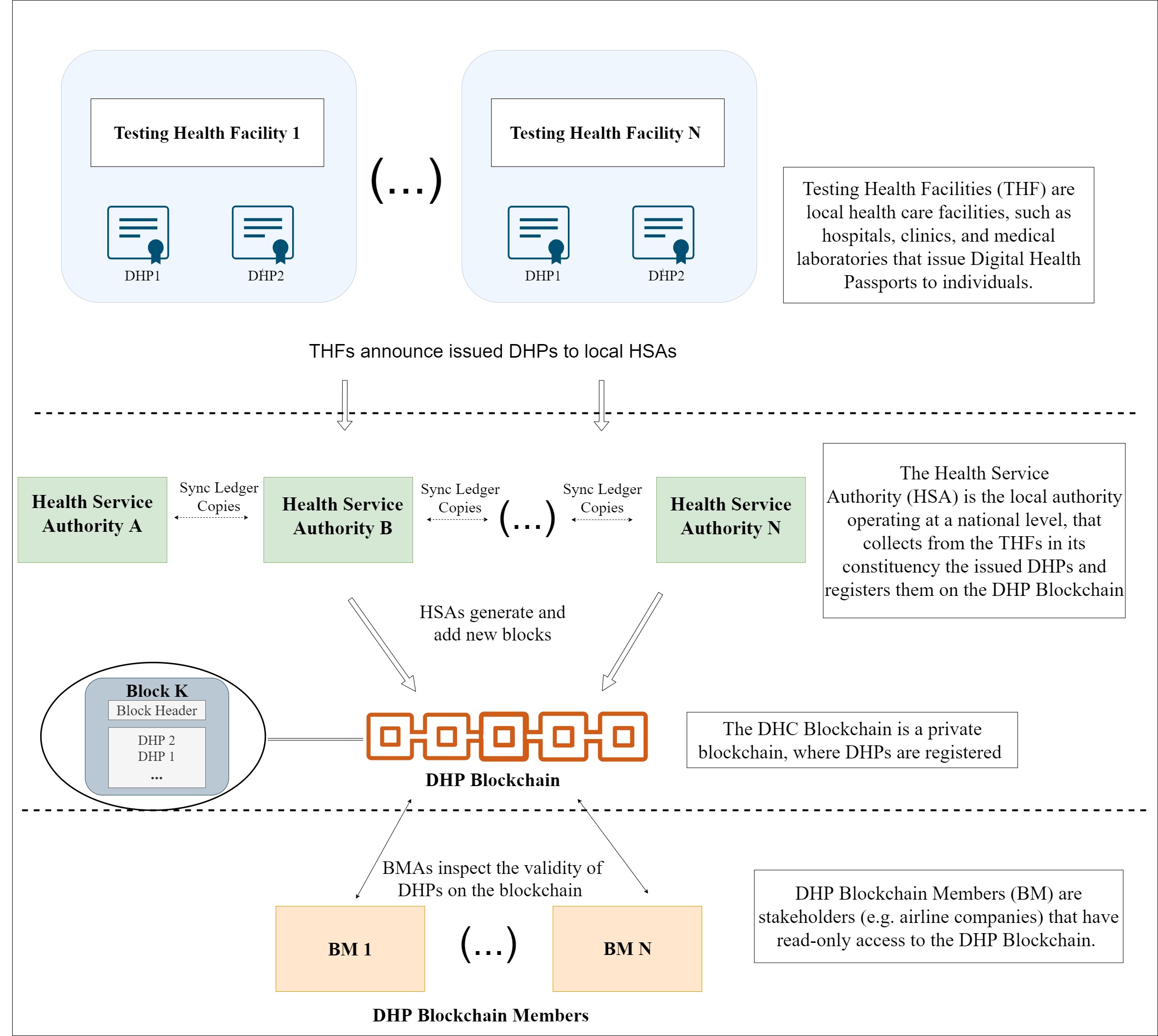}
\caption{Reference architecture of the DHP Framework.}
\label{fig:architecture}
\vspace{-0.5cm}
\end{figure*}

\section{The Digital Health Passports Framework}
\label{framework}
\subsection{Actors and data structures in the DHP Framework.}
We identify the following actors and data structures:

\textbf{Testing Health Facilities (THF)} are local health care facilities - such as hospitals, clinics, and medical laboratories - where individuals can get tested on whether they are active carriers of the virus (i.e. if they can transmit the virus). THFs are the entities responsible for verifying the identity of the individuals and the validity of the test results. THFs would participate in a public key infrastructure (PKI) scheme to produce and sign the digital passports. In essence, the health certificate can be expressed as a conventional X.509 type of digital passport with an appropriate scheme capturing the information that is appropriate in this problem domain. The use of a digital passport will enable the party that is responsible for publishing the results on the blockchain to authenticate the information produced and sent by the THF. The digital passport scheme will however need to contain information in the form to support the security and privacy properties as described later on.

The technical details of the test, such as the time it takes to produce the results, its accuracy and precision, its type, etc., are considered external parameters specified by corresponding medical experts and are out of the scope of the presented system. We note, that regarding SARS-CoV-2 and the associated disease, COVID-19, at the time of writing there is no scientific evidence that people who have recovered from COVID-19 and have antibodies are protected from a second infection \cite{whopassport}. Therefore, we consider that the results of a test are valid for a specified period of time beyond which the person needs to be tested again. We also anticipate that over time new testing methods will be developed and therefore the type of test used is a parameter that needs to be captured.

We define the \textbf{Digital Health Passport (DHP)} for individual $i$ at time $j$ as
\begin{equation}
\textbf{c}_{i,j} = \langle p_{i}; r_j; t_j; \sigma; s_{THF} \rangle
\label{def:dhc}
\end{equation}
where $p_{i}$ is the output of a privacy preserving cryptographic primitive\footnote{We elaborate on the role of this field in Section \ref{sec:privacy-analysis}}; $r_i$ is the boolean result of the test (i.e., the individual is risk free or not); $t_i$ is a timestamp specifying the date of the result; $\sigma$ denotes the testing method used; and $s_{THF}$ denotes the digital signature of the THF for this certificate.  

The \textbf{DHP Blockchain} is a \textit{private} blockchain where DHPs are registered. As a private blockchain it does not allow public access to the data it contains and the identity of all participating entities is verified and well-known. The participating entities are of two types (defined in the next paragraphs); (a) the Health Service Authorities that have full access rights on the blockchain; and (b) the Blockchain Members  that have only read rights on the blockchain. In the DHP Blockchain the Health Service Authorities use \textit{Proof of Authority} as a consensus mechanism \cite{network2017proof}.  

The \textbf{Health Service Authority (HSA)} is the local authority, operating at a national level (e.g. the NHS in the UK), that collects from the THFs in its constituency the issued DHPs and registers them on the DHP Blockchain. This is considered a trustworthy and honest entity that maintains and grows the DHP Blockchain in collaboration with its peer HSAs from other countries. In blockchain terms, HSAs batch multiple DHPs together and generate the corresponding new blocks that are appended to the DHP Blockchain. HSAs are the only entities that have the right to generate and append new blocks. 

The \textbf{DHP Blockchain Member (BM)} is an entity that is member of the DHP Blockchain consortium but only has read access rights. This allows the BM to check if a person is the holder of a valid DHP. Such entities include airline companies, airport security, border control authorities, and other relevant stakeholders.

\subsection{Procedure Description}

In order for an individual to make an international travel, they will need to be the holders of a valid travel document as well as of a valid Digital Health Passport. It is also assumed that they have an adequately secure smartphone device to participate in the verification. Unlike the digital contact tracing apps that introduce risks to the individual's privacy, the proposed app is used to enforce the privacy of the individual's result. In the case of the app being compromised (assuming that the privacy preserving protocol computations fail), the user will suffer a Denial of Service with respect to using the DHP Framework. Mitigations of this type of risks are outside the scope of this paper. Moreover, it should be noted that the proposed approach can also work without the use of a smartphone and app, but it should be noted that in this case the problem is reduced to a plain verification of an individual's examination results without the privacy preservation aspects. In this case, $p_i$ would be reduced to a straightforward cryptographic one-way hash function.

The process is structured in two phases - the \textit{DHP issuance }and the \textit{DHP verification}, as follows.\newline

\textbf{Phase 0:} \textit{Citizen registration}	
\begin{enumerate}[align=left]
\item[\textit{Step 1:}] The individual downloads and registers on the DHP app. The registration involves the generation of a type of digital wallet where the citizen's travel document - along with the accompanying parameters set out by the underlying encryption scheme - is encrypted by their public key. This ensures that the citizen can prove that they are the individuals referenced by the respective travel document. 
\end{enumerate}

\textbf{Phase 1:} \textit{Digital Health Passport issuance}	
\begin{enumerate}[align=left]
\item[\textit{Step 1:}] The individual arrives at their local THF and declares that they would like to be issued a DHP for international travel. They present their travel document (out-of-band) and get tested for the virus.
\item[\textit{Step 2:}] When the results are ready, the traveller is informed of the outcome. If the results indicate that they are risk free, they request from the THF to issue a DHP; by doing so, they also provide their consent for the THF and the corresponding HSA to register the DHP on the DHP Blockchain.
\item[\textit{Step 3:}] The THF issues the DHP corresponding to the results of that particular test using the travel document to generate the DHP parameters. The THF then transmits the DHP to the corresponding HSA.
\item[\textit{Step 4:}] The HSA collects the DHPs issued by the THFs in its constituency and proceeds with generating the corresponding blocks and appending them to the DHP Blockchain.
\item[\textit{Step 5:}] The HSA informs the THFs of the block headers for each registered DHP they issued. In turn, the THF provides the individual with a DHP token containing the block header corresponding to their new DHP. \newline
\end{enumerate}  

\textbf{Phase 2:} \textit{Digital Health Passport verification}
\begin{enumerate}[align=left]
\item[\textit{Step 1:}] The individual proceeds with purchasing their air flight ticket. During their check-in, they provide the airline company with their travel document, their DHP token and they give consent through the DHP app for the airline company to access their examination results record instance. The latter triggers a process that sends the required data to perform searchable encryption operations on the DHP Blockchain. 
\item[\textit{Step 2:}] The airline (a BM with read-only rights on the DHP Blockchain) uses the DHP token to identify and retrieve the DHP in the corresponding block. The airline performs searchable encryption (or a hash lookup in the case of the citizen opting out of the app) of the travel document and identifies the corresponding health passport in the retrieved block.  
\item[\textit{Step 3:}] The airline verifies the integrity of the DHP using its digital signature. It then proceeds with verifying the validity of the DHP; i.e. if the test result, timestamp and test type adhere to the the hygiene protocol enforced by the destination authorities.    
\item[\textit{Step 4:}] If the certificate adheres to the hygiene protocol in force at the destination, the boarding pass is issued and the individual proceeds with their travel. 
\end{enumerate}

\section{DHP Framework Analysis}
\label{analysis}

\subsection{Usability and efficiency}
The use case we consider on international tourism  involves the two heavily regulated sectors of health and airline industry. This facilitates the implementation of the presented DHP Framework, since the identity of all involved actors will be easily validated, therefore, contributing to their accountability. Furthermore, both sectors are experienced with protocols and processes of identity validation, and storage and management of personal data. This contributes to the positive public perception and acceptance of the system, particularly when compared to the concerns raised with regards to DCT applications (consider the act of visiting a medical facility versus installing a mobile application that explicitly collects location or social interaction data). The framework makes use of the Proof of Authority consensus mechanism, which allows for a short transaction time (i.e. the amount of time needed for new data to be registered on the blockchain) with low demands on computing power, therefore making the operation of the framework cost effective.  

The DHP Framework makes use of a distributed ledger where the validators of new blocks are the national health authorities of each country. This allows for countries to participate in the digital health passport scheme in a controlled and scalable way, while at the same time obfuscating any heterogeneities. For instance, national health systems may use different diagnostic methods, may have different underlying structures, or may enforce different hygiene protocols that airline companies and border control authorities need to follow for allowing incoming international travellers. The presented DHP Framework provisions the registration of all needed data that is needed by the stakeholders for this purpose and also provides a tools for seamless, continuous, secure and timely exchange of such data.   

\subsection{Security considerations}
Following, we elaborate on the security aspects of the DHP Framework.
\subsubsection{Integrity} Each DHP is digitally signed by the issuing THF using asymmetric encryption, and therefore its integrity is guaranteed. Furthermore, each DHP is linked to a specific person via the corresponding travel document and therefore is not transferable to another person.

\subsubsection{Attribution and accountability} The issuing THF of each DHP can be identified by its digital signature. Therefore, each individual THF is accountable for issuing true and correct DHPs and can be back-tracked if needed (e.g. in the case of a THF issuing fraudulent DHPs).

\subsubsection{Immutability} The DHP Framework provides guarantees on the immutability of the record of the registered DHPs, as derived from the use of blockchain. This is in contrast to a physical health passport (e.g. in the form of a document) that can be subject to forgery.  

\subsection{Privacy considerations}
\label{sec:privacy-analysis}
Following we elaborate on the privacy aspects of the DHP Framework.
\subsubsection{Anonymity} The liaison between a DHP and an individual is the corresponding travel document which is  stored using a privacy preserving cryptographic
primitive. Therefore, one can not retrieve the details of the DHP holder, unless they are provided with the details of the travel document. 

\subsubsection{Right to be forgotten} The individual explicitly consents to the DHP being registered on the DHP Blockchain when initiating the process at the THF. Then the individual proceeds with providing the airline company with their travel document during chekc-in. Therefore, the airline company can now parse the distributed ledger in order to retrieve additional entries corresponding to the same individual. Although this does not apply globally, in several regions there is legislation in place that safeguards the citizens' right to be forgotten (e.g. the General Data Protection Regulation in the European Union) and the individual can request that all records including personal data are deleted. 

\section{Future Considerations}
\label{future}
In this section we identify formal security and privacy properties that need to be guaranteed rigorously in future research. Developing corresponding security protocols is the subject of future research.

\subsection{Security properties}

Considering that the DHP Framework is based on the deployment of a distributed ledger, we argue that the following properties need to be met:

\subsubsection{$\theta$-Liveness}
All reporting HSAs will report honestly submitted items within some delay bound $\theta$ \cite{Kiayias2018}. 

\subsubsection{Consistency}
At any given point in time, all verifying parties should obtain the same outcome. As a traveller may be tested more than once, there should be no potentially conflicting outcomes. This property, combined with $\theta$-liveness ensures that the information on the examination result will be valid and timely.

\subsubsection{Verifiability}
It should be possible for a BM to associate a time bounded examination result with a given traveller (currently achieved in the DHP Framework using the DHP timestamp).

\subsubsection{Attribution}
It should be possible to identify the THF of a particular test result. This is necessary, as the party performing the test can be different from the party reporting and publishing the result (currently achieved in the DHP Framework by the digital signature of each THF).

\subsubsection{Unforgeability}
A DHP should be unforgeable. This can be achieved with the use of a public key infrastructure (currently addressed in the DHP Framework by the digital signature of each THF).

\subsubsection{Accountable verifiability}
It should be possible for a BM to demonstrate that they performed a verification process for all travellers included in a given manifest. We conjecture that this property can be offered through the deployment of smart contracts.

\subsection{Privacy properties}
\label{sec:privacy-analysis}

\subsubsection{Unlinkability}
It should not be possible for a BM to associate all past exams recorded on the ledger with a particular traveller; The BM should have restricted and local scope that is bounded by the time applicable to a particular travel event.

\subsubsection{Unexplorability}
It should not be possible for a curious verifier to traverse the blockchain and obtain examination information from individuals who are not in the manifest of a particular travel event.

\subsubsection{Anonymity}
All data pertaining to a citizen's health examination results should not be readily linked to an individual.

\balance
\bibliographystyle{IEEEtran}
\bibliography{covid19}

\end{document}